\documentclass[%
superscriptaddress,
prb,
reprint,
 amsmath,amssymb,
]{revtex4-1}

\usepackage{graphicx}%
\usepackage{dcolumn}%
\usepackage{bm}%
\usepackage[mathlines]{lineno}%
\setcitestyle{super}

\usepackage{color}

\begin{document}

\title{Polarization-state-resolved high-harmonic spectroscopy of solids}

\author{N. Klemke}
\thanks{These two authors contributed equally}
\affiliation{Center for Free-Electron Laser Science CFEL, \\Deutsches Elektronen-Synchrotron DESY, Notkestra\ss e 85, 22607 Hamburg, Germany}
\affiliation{Physics Department, University of Hamburg, Luruper Chaussee 149, 22761 Hamburg, Germany}

\author{N. Tancogne-Dejean}
\thanks{These two authors contributed equally}
\email{nicolas.tancogne-dejean@mpsd.mpg.de}
\affiliation{Center for Free-Electron Laser Science CFEL, \\Deutsches Elektronen-Synchrotron DESY, Notkestra\ss e 85, 22607 Hamburg, Germany}
\affiliation{Max Planck Institute for the Structure and Dynamics of Matter, Luruper Chaussee 149, 22761 Hamburg, Germany}

\author{G. M. Rossi}
\affiliation{Center for Free-Electron Laser Science CFEL, \\Deutsches Elektronen-Synchrotron DESY, Notkestra\ss e 85, 22607 Hamburg, Germany}
\affiliation{Physics Department, University of Hamburg, Luruper Chaussee 149, 22761 Hamburg, Germany}

\author{Y. Yang}
\affiliation{Center for Free-Electron Laser Science CFEL, \\Deutsches Elektronen-Synchrotron DESY, Notkestra\ss e 85, 22607 Hamburg, Germany}
\affiliation{Physics Department, University of Hamburg, Luruper Chaussee 149, 22761 Hamburg, Germany}

\author{R. E. Mainz}
\affiliation{Center for Free-Electron Laser Science CFEL, \\Deutsches Elektronen-Synchrotron DESY, Notkestra\ss e 85, 22607 Hamburg, Germany}
\affiliation{Physics Department, University of Hamburg, Luruper Chaussee 149, 22761 Hamburg, Germany}

\author{G. Di Sciacca}
\affiliation{Center for Free-Electron Laser Science CFEL, \\Deutsches Elektronen-Synchrotron DESY, Notkestra\ss e 85, 22607 Hamburg, Germany}

\author{A. Rubio}
\email{angel.rubio@mpsd.mpg.de}
\affiliation{Center for Free-Electron Laser Science CFEL, \\Deutsches Elektronen-Synchrotron DESY, Notkestra\ss e 85, 22607 Hamburg, Germany}
\affiliation{Physics Department, University of Hamburg, Luruper Chaussee 149, 22761 Hamburg, Germany}
\affiliation{Max Planck Institute for the Structure and Dynamics of Matter, Luruper Chaussee 149, 22761 Hamburg, Germany}
\affiliation{The Hamburg Centre for Ultrafast Imaging, Luruper Chaussee 149, 22761 Hamburg, Germany}
\affiliation{Center for Computational Quantum Physics (CCQ), The Flatiron Institute, 162 Fifth Avenue, New York NY 10010, USA}

\author{F. X. K\"artner}
\affiliation{Center for Free-Electron Laser Science CFEL, \\Deutsches Elektronen-Synchrotron DESY, Notkestra\ss e 85, 22607 Hamburg, Germany}
\affiliation{Physics Department, University of Hamburg, Luruper Chaussee 149, 22761 Hamburg, Germany}
\affiliation{The Hamburg Centre for Ultrafast Imaging, Luruper Chaussee 149, 22761 Hamburg, Germany}

\author{O. D. M\"ucke}
\email{oliver.muecke@cfel.de}
\affiliation{Center for Free-Electron Laser Science CFEL, \\Deutsches Elektronen-Synchrotron DESY, Notkestra\ss e 85, 22607 Hamburg, Germany}
\affiliation{The Hamburg Centre for Ultrafast Imaging, Luruper Chaussee 149, 22761 Hamburg, Germany}

\date{\today}%

\begin{abstract}
\noindent Attosecond metrology \cite{KrauszIvanovRMP} sensitive to sub-optical-cycle electronic and structural dynamics is opening up new avenues for ultrafast spectroscopy of condensed matter.
Using intense lightwaves to precisely control the extremely fast carrier dynamics in crystals holds great promise for next-generation electronics and devices operating at petahertz bandwidth \cite{KrauszStockman_NP14,Mashiko_NP16}. The carrier dynamics can produce high-order harmonics of the driving light field extending up into the extreme-ultraviolet region \cite{Ghimire_NP11,Luu_Nat15,Garg_Nat16}.
Here, we introduce polarization-state-resolved high-harmonic spectroscopy of solids, which provides deeper insights into both electronic and structural dynamics occuring within a single cycle of light.
Performing high-harmonic generation measurements from silicon and quartz samples, we demonstrate that the polarization states of high-order harmonics emitted from solids \cite{Ghimire_NP11} are not only determined by crystal symmetries, but can be dynamically controlled, as a consequence of the intertwined interband and intraband electronic dynamics
\cite{Huttner_LPR17,nicolasPRL,nicolasncomm17}, responsible for the harmonic generation.
We exploit this symmetry-dynamics duality to efficiently generate circularly polarized harmonics from elliptically polarized driver pulses. Our experimental results are supported by \textit{ab-initio} simulations \cite{nicolasPRL,nicolasncomm17}, providing clear evidence for the microscopic origin of the phenomenon.
This spectroscopy technique might find important applications in future studies of novel quantum materials \cite{BasovNM17} such as strongly correlated materials.
Compact sources of bright circularly polarized harmonics in the extreme-ultraviolet regime will advance our tools for the spectroscopy of chiral systems, magnetic materials, and 2D materials with valley selectivity.
\end{abstract}

\maketitle

The study of lightwave-driven electronic dynamics occurring on sub-optical-cycle time scales in condensed matter and nanosystems is a fascinating frontier of attosecond science originally developed in atoms and molecules \cite{KrauszIvanovRMP,CiappinaRPP}.
Adapting attosecond metrology techniques to observe and control the fastest electronic dynamics in the plethora of known solids and novel quantum materials \cite{BasovNM17} is very promising for studying correlated electronic dynamics (e.g., excitonic effects, screening) on atomic length and time scales,
thereby potentially impacting future technologies such as emerging petahertz electronic signal processing \cite{KrauszStockman_NP14,Mashiko_NP16} or strong-field optoelectronics
\cite{Schiffrin_Sci13,Sivis_Sci17}.

The nonlinear process of high-order harmonic generation (HHG) in gases is one of the cornerstones of attosecond science and is well understood by the semiclassical three-step model \cite{KrauszIvanovRMP}. In solids, nonperturbative HHG up to 25th harmonic order without irreversible damage was first reported in [\onlinecite{Ghimire_NP11}].
This work triggered extensive research activities aimed at unraveling the microscopic interband and intraband dynamics underlying HHG from crystals (for a comprehensive overview, see [\onlinecite{Huttner_LPR17}]), thereby extending attoscience techniques to solids. The prevailing strong-field dynamics were successfully identified in specific cases, even if a global picture has not yet emerged. Other works demonstrated isolated attosecond extreme ultraviolet (XUV) pulses emitted from thin SiO$_2$ films \cite{Luu_Nat15,Garg_Nat16}, or investigated HHG from
2D materials such as graphene \cite{Yoshikawa_Sci17}, 2D transition-metal dichalcogenides \cite{Yoshikawa_Sci17,Liu_NP17}, and monolayer hexagonal boron nitride \cite{Nicolas2D}.

Elucidating the complex microscopic electronic dynamics producing HHG without making \emph{a-priori} severe assumptions poses a challenge for theory. Indeed, the theory must capture at the same time the transitions between discrete electronic bands, and the ultrafast motion of electrons within the bands; two mechanisms usually decoupled in the description of either optical properties or transport in semiconductors and insulators. An effective way to account for the full interacting many-body electronic dynamics and real crystal structure is using \emph{ab-initio} time-dependent density-functional theory (TDDFT) simulations  \cite{nicolasPRL,nicolasncomm17}. Some of us recently used this theoretical framework to reveal how the microscopic mechanisms governing HHG in solids depend on the ellipticity of the driving field and the underlying band structure \cite{nicolasncomm17}. That work predicted that different harmonics react differently to the driver ellipticity, as they can either originate mainly from intraband contributions or from coupled interband and intraband dynamics \cite{nicolasPRL}.

The symmetry properties of the light-matter interaction Hamiltonian distinguishes HHG in crystals from atoms and molecules, with major ramifications for the selection rules of different harmonics and their polarization states.
HHG from atoms driven by propeller-shaped bichromatic waveforms produces circular (see remark on terminology \cite{linearcircularelliptical}) harmonics
\cite{Kfir2014}.
In molecules, both the point group
and the driving field determine the symmetries of the coupled light-matter system. Consequently, depending on the molecular symmetries and the molecule's orientation with respect to the light propagation direction, elliptical high-order harmonics can be produced by linear
\cite{Smirnova09,Mairesse10} or elliptical driver pulses \cite{Cireasa15}.
For bichromatic bicircular driver fields, circular harmonics with alternating helicities can be obtained, provided that the molecule's symmetries are compatible with that of the driving field
\cite{Bandrauk16,Reich16,Baykusheva16}.
In crystals, several recent works studied the high-harmonic response on driver pulse ellipticity $\epsilon$, which can strongly differ qualitatively from the atomic and molecular cases. Whereas earlier work \cite{You_NP16} looked exclusively at the harmonic yield, later research also investigated the polarization states and selection rules of the higher harmonics from various solids of different crystal symmetries \cite{nicolasncomm17,Klemke_Atto17,Saito_Optica17} and reported \emph{circular} HHG from a single-color driver field \cite{Klemke_Atto17,Saito_Optica17}, which is symmetry forbidden in atoms.

Here, we present a combination of HHG experiments and first-principles TDDFT simulations for silicon and quartz, demonstrating that a complete
understanding of the harmonics' polarization states requires, beside knowledge of the crystal's symmetries, a microscopic understanding of the underlying complex, coupled interband and intraband dynamics \cite{nicolasPRL,nicolasncomm17}. Most importantly, we demonstrate for the first time the strong-field control of the harmonics' polarization states.
Our findings indicate that polarization-state-resolved high-harmonic spectroscopy of solids provides deeper insights into both electronic and structural dynamics as well as symmetries on sub-cycle time scales.

In our experiments, we irradiated free-standing, 2-$\mu$m-thin, (100)-cut silicon samples with 120 fs, 2.1 $\mu$m (0.59 eV) pulses with tunable ellipticity $\epsilon$ and peak intensities up to 0.7\,TW\,cm$^{-2}$ in vacuum (see Methods section and Supplementary Fig. 1). At this intensity, the harmonics are generated nonperturbatively (see Supplementary Fig. 3) up to harmonic order 19 (HH19) in the XUV regime for our experimental conditions, as shown by our TDDFT simulations (see Supplementary Fig. 5). Only harmonics up to HH9 are detected by the spectrometer used in our experiments. We also irradiated 50-$\mu$m-thin, z-cut quartz with an estimated intensity of 40\,TW\,cm$^{-2}$ in vacuum.

\begin{figure}[t]
    \centering
    \includegraphics[width=\columnwidth]{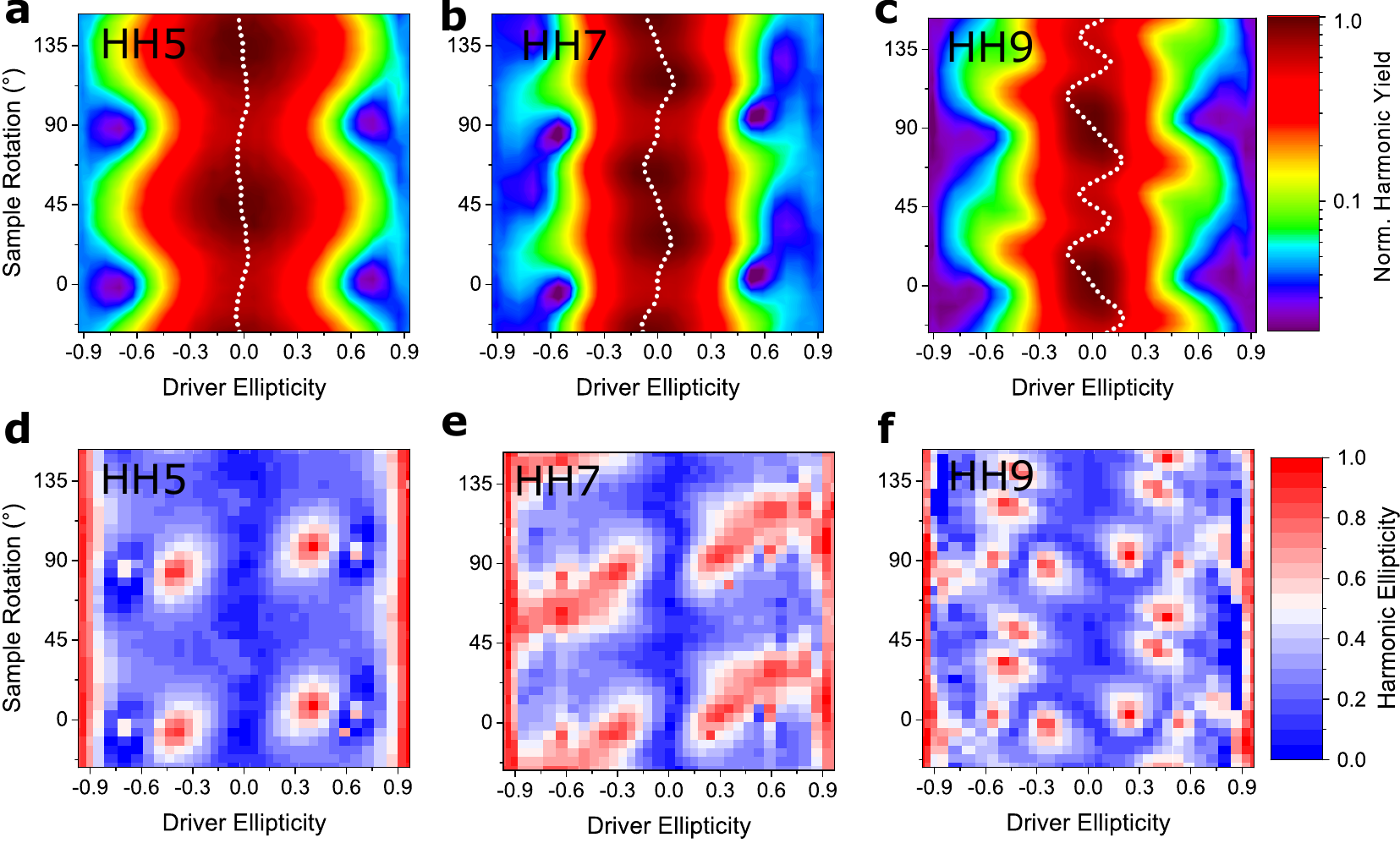}
    \caption{\label{maps}\textbf{High-harmonic response of silicon versus driving pulse ellipticity and sample rotation.}
    Measured intensity and harmonic ellipticity $|\epsilon_{\rm HH}|$ of HH5 (\textbf{a}, \textbf{d}), HH7 (\textbf{b}, \textbf{e}) and HH9 (\textbf{c}, \textbf{f}) as function of driver ellipticity and sample rotation. The white dotted lines in (\textbf{a})-(\textbf{c}) indicate the centers of mass ($\times$5 to enhance visibility of the variation) of the intensity distributions. $0^{\circ}$ and $90^{\circ}$ sample rotation correspond to driver major axis along $\Gamma$X, $45^{\circ}$ and $135^{\circ}$ along $\Gamma$K. The peak driving intensity is 0.6\,TW\,cm$^{-2}$ in vacuum.}
\end{figure}

Fig. \ref{maps} shows the measured high-harmonic response in Si of HH5, HH7 and HH9
as function of driver ellipticity $\epsilon$ and sample orientation $\theta$; panels a-c display normalized harmonic intensities, d-f harmonic ellipticities $|\epsilon_{\rm HH}|=\sqrt{I_{\mathrm{min}}/I_{\mathrm{max}}}$, where $I_{\mathrm{min}}$ and $I_{\mathrm{max}}$ correspond to the intensities at the minor and major axis of the polarization ellipse.
In all panels, $\theta = 0^{\circ}$, $45^{\circ}$ refer to the major axis of the driving field ellipse along the directions [100] ($\Gamma$X) and [110] ($\Gamma$K) in real (reciprocal) space.
The crystal symmetries are recovered in all maps shown in Fig. \ref{maps}.

All harmonics respond in a distinctly different way to the driver-pulse ellipticity $\epsilon$, and the harmonic yields peak for different sample rotations. HH5 exhibits Gaussian-shaped, atomic-like ellipticity profiles for all sample rotations.
The distribution is symmetric around $\epsilon = 0$ (see white dotted center-of-mass curve) for all sample rotations.
The intensity distribution of HH7 (Fig. \ref{maps}b) shows intriguing, non-atomic-like features with maximum yield at non-zero ellipticity for certain sample rotations, similar to experiments on MgO.\cite{You_NP16}
HH9 (Fig. \ref{maps}c) exhibits the most pronounced deviations from a Gaussian-like ellipticity profile, with non-monotonic non-atomic-like profiles for wide ranges of sample rotation.
Its yield is strongly asymmetric with respect to $\epsilon = 0$ for all sample orientations (different from mirror planes), and displays strong non-sinusoidal oscillations of the center-of-mass curve.

The overall behavior can be understood by inspecting the Si band structure:
HH5 (2.95 eV) is below the direct Si bandgap of 3.1 eV. This harmonic thus originates purely from intraband dynamics of low-energetic electrons, that mostly remain within the parabolic region of the bands, leading to an atomic-like behavior.
For above-bandgap harmonics, the joint density of states (JDOS) (see Supplementary Fig. 6), i.e., the density of optical transitions at a given energy, determines the relative weight of interband compared to intraband mechanisms \cite{nicolasPRL}. Around 5.3 eV (HH9), the JDOS is significantly lower than for 4.1 eV (HH7).
Therefore, while coupled inter- and intraband dynamics lead to the emission of HH7, HH9 is mostly produced by intraband effects \cite{nicolasPRL}. Interestingly, these harmonics are more efficiently generated with different helicities, as can be seen from the different signs of the center-of-mass curves for certain sample rotations (see Supplementary Fig. 7). This clearly indicates different generation mechanisms of HH7 and HH9, as predicted in Ref. [\onlinecite{nicolasncomm17}].
For HH9, for which interband transitions are strongly suppressed by the low JDOS at this energy, higher-energetic electrons explore larger non-parabolic regions in the bands, which results in pronounced non-atomic-like ellipticity profiles.

\begin{figure}[b]
    \centering
    \includegraphics[width=\columnwidth]{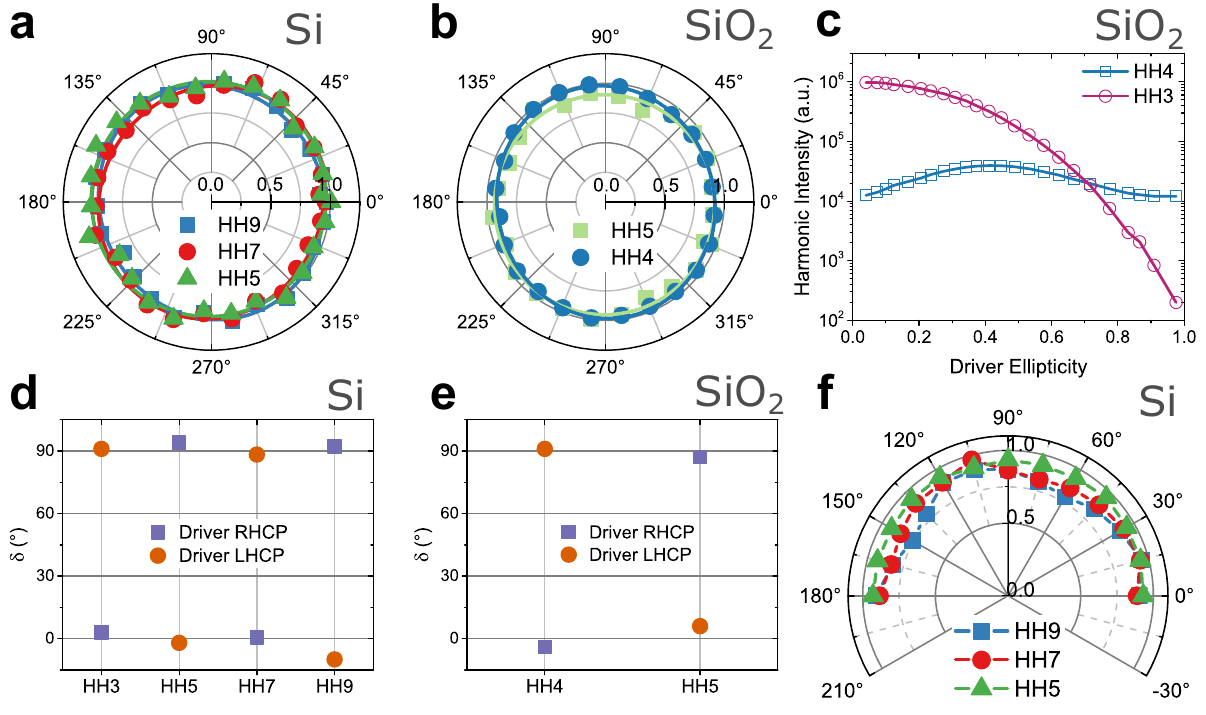}
    \caption{\label{circdriver}\textbf{Measured circular harmonics from a circular driver and selection rules.} Normalized harmonic intensity versus polarizer rotation angle from silicon (\textbf{a}) and quartz (\textbf{b}), showing circular harmonics. The solid lines are sin-square fits. (\textbf{c}) Intensity of HH3 and HH4 from quartz versus driver ellipticity.
     Harmonic major-axis rotation after a second quarter-wave plate for silicon (\textbf{d}) and quartz (\textbf{e}), indicating alternating helicities (RHCP/LHCP, right/left handed circular polarization) with harmonic order, consistent with selection rules. (\textbf{f}) Harmonic ellipticities $|\epsilon_{\rm HH}|$ from Si versus sample rotation.}
\end{figure}

Fig. \ref{maps}d-f reports the measured harmonics' polarization states as function of driving ellipticity and sample rotation.
Whereas linear drivers yield almost linear harmonics, we observe astonishing deviations of the harmonic ellipticities $\epsilon_{\rm HH}$ from the driver ellipticity $\epsilon$.
Consistent with our TDDFT predictions \cite{nicolasncomm17} and selection rules in [\onlinecite{Tang71}], for circular driver pulses, $|\epsilon| \approx 1$, all harmonics become circular $|\epsilon_{\rm HH}| \approx 1$. Most importantly, for all observed harmonics, circular harmonics can be generated from elliptical driving polarizations, as elaborated on below.
These 'islands' of high ellipticity sensitively depend on $\epsilon$ and $\theta$ in the cases of HH5 and HH9, however, for HH7 this sensitivity is less pronounced.
This observation is again consistent with a strong dependence of the microscopic mechanisms on the polarization state of the driving field, as the electrons explore different regions of the BZ depending on $\epsilon$ and $\theta$. The measured harmonics' polarizations contain the complete information on the $x$- and $y$-components of the harmonics' amplitudes and their relative phases.

Fig. \ref{circdriver} summarizes our findings on circular harmonics from circular drivers.
In both silicon (Fig. \ref{circdriver}a) and $\alpha$-quartz (Fig. \ref{circdriver}b), all harmonic intensities remain constant while rotating a polarizer by $360^{\circ}$, thus
confirming circular harmonic polarization.
In Fig. \ref{circdriver}c, we observe a strong intensity suppression of HH3 going from linear to circular driver, as expected from the selection rules for the $D_3$ [$32$] point group \cite{Tang71} of $\alpha$-quartz.
The selection rules also manifest themselves in the helicities of the circular harmonics. In accordance with group-theoretical considerations \cite{Tang71} and TDDFT simulations \cite{nicolasncomm17}, the odd harmonics from Si have alternating helicities as Si has point group $O_h$ [$m3m$].
This was confirmed with a tunable quarter-wave plate behind the sample, which converts circular to linear polarization, with the polarization angle $\delta$ depending on the helicity (see Fig. \ref{circdriver}d). The trigonal crystal structure of $\alpha$-quartz results in different selection rules, leading to alternating helicities of HH4 and HH5 in Fig. \ref{circdriver}e.
As shown in Section VIII of the Supplementary Information, we extracted the Stokes polarization parameters of the harmonics from these measurements and estimated a value of the degree of polarization (DOP) of $0.8 \pm 0.2$ for all harmonics, similar to reported values for the generation of circular harmonics from atomic and molecular gases \cite{Kfir2014,Veyrinas16}.
Moreover, we find in Si that the harmonic ellipticities $|\epsilon_{\rm HH}|$ are all close to 1, independent of sample rotation $\theta$ (see Fig.~\ref{circdriver}f).
Minor deviations from $|\epsilon_{\rm HH}|=1$ are likely due to small deviations from perfectly circular driver pulses.

\begin{figure}[b]
    \centering
    \includegraphics[width=\columnwidth]{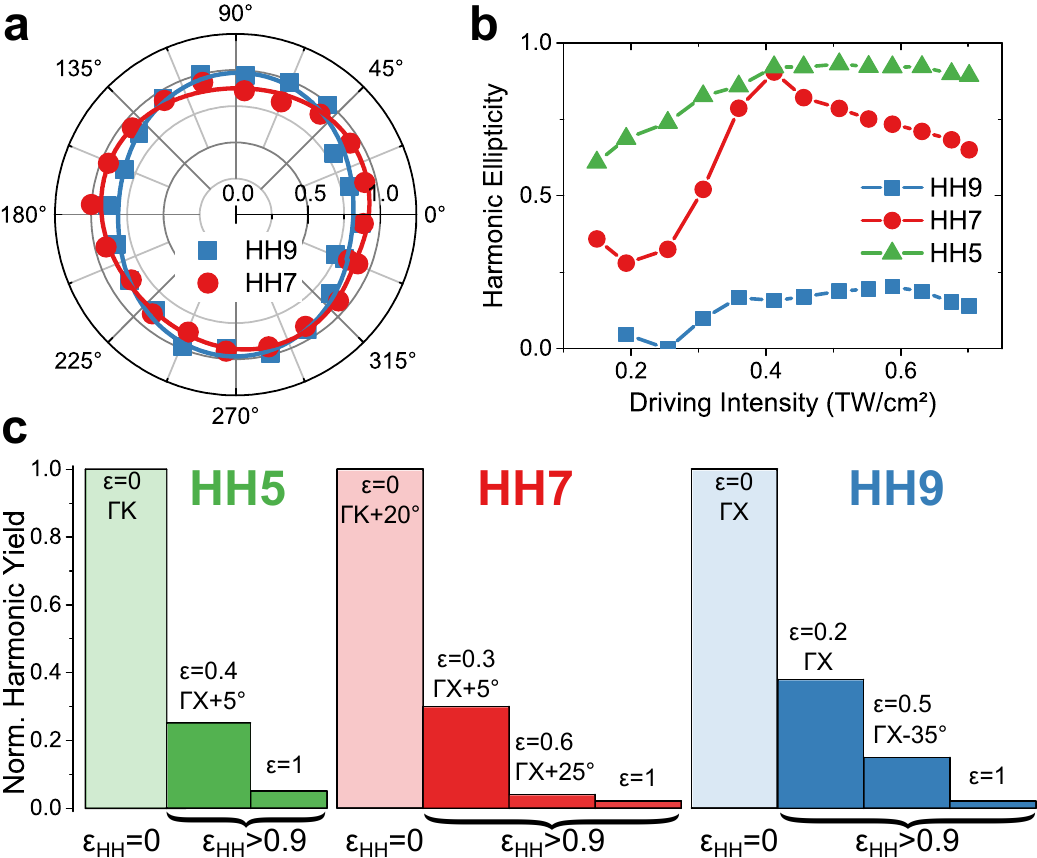}
    \caption{\label{ellipdriver}\textbf{Measured circular harmonics from elliptical driver pulses in Si.} (\textbf{a})
    Polarizer scan of HH9 ($\epsilon = 0.5$, $\theta = \Gamma\text{X}+30^{\circ}$) and HH7 ($\epsilon = 0.3$, $\theta = \Gamma\text{X}+5^{\circ}$). The solid lines are sin-square fits. (\textbf{b}) Harmonic ellipticities versus driving intensity for $\epsilon = 0.4$ and $\theta = \Gamma\text{X}+10^{\circ}$. (\textbf{c}) Yields of HH5-HH9 for exemplary cases of circular harmonic polarization. The harmonic yields are normalized to the maximum harmonic yield for $\epsilon = 0$ (indicated by the light color bars). }
\end{figure}

Fig. \ref{ellipdriver}a shows two polarizer scans under excitation conditions, for which HH9 and HH7 are circular for elliptical driver polarization. The measured harmonic ellipticities are $\sim$0.93 in both cases. We also found similarly high ellipticities for HH5 (not shown).
Fig. \ref{ellipdriver}b shows the intensity-dependence of the harmonic ellipticities for $\epsilon = 0.4$ and $\theta = \Gamma\text{X} + 10^{\circ}$.
By varying the intensity of the driving field, we achieve a high degree of control over the harmonics' polarization states. This key result has two important consequences:
First, it shows that the relative importance of interband and intraband mechanisms is not a material property only, but strongly depends on excitation conditions, thus offering a broader perspective on the controversial debate about the dominant mechanism responsible for HHG in solids.
Second, the observation of circular harmonics for elliptical driver polarization, that sensitively depend on the nonperturbative dynamics of the system, can \emph{not} be explained by symmetry arguments only, but clearly indicates strong-field control of the harmonic ellipticities $|\epsilon_{\rm HH}|$ through the lightwave-driven electron dynamics. This might find applications, e.g., in polarization-controlled high-harmonic sources.

The total harmonic intensities for exemplary cases of circular harmonics for different $\epsilon$ and $\theta$ are compared in Fig. \ref{ellipdriver}c.
As discussed above, for Si, the harmonic yield tends to decrease (apart from non-monotonic exceptions) with increasing $|\epsilon|$. Therefore, the generation of circular harmonics using elliptical driver pulses ($|\epsilon|<1$) is expected to be significantly more efficient than for circular ones ($|\epsilon|=1$), as indeed observed in Fig. \ref{ellipdriver}c.
For HH5 and HH7, the circular harmonics generated for $\epsilon = 0.3-0.4$ and $\theta = \Gamma\text{X}+5^{\circ}$ are 10$\times$ brighter than for circular driver pulses.
In the case of HH9, circular harmonics were even produced with 40\% efficiency compared to maximum yield obtained for linear polarization, which corresponds to an 18$\times$ yield enhancement going from $\epsilon = 1$ to $\epsilon =0.2$.

\begin{figure}[t]
    \centering
    \includegraphics[width=\columnwidth]{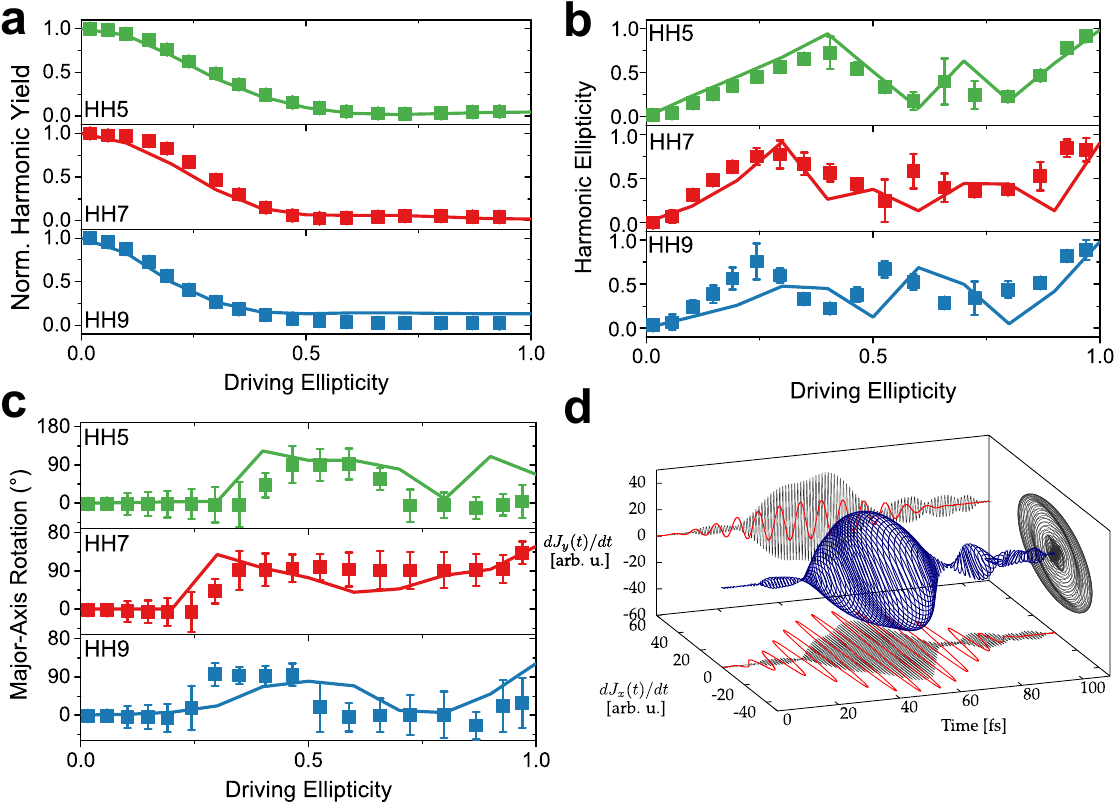}
    \caption{\label{CompExpTDDFT}\textbf{Calculated polarization states of the harmonics compared to the experiment.} Comparison between TDDFT simulations (solid lines) and experimental results for (\textbf{a}) the harmonic yield, (\textbf{b}) the harmonic ellipticity,  and (\textbf{c}) the major-axis rotation of the harmonics' polarization ellipse of HH5 to HH9 versus the driving ellipticity. Here, $\theta = \Gamma\text{X}$. For all plots, the values are interpolated between $\theta = \Gamma\text{X}^{+2^{\circ}}_{-3^{\circ}}$ and averaged over negative and positive ellipticity values. The error bars are the averaged absolute deviations.
    (\textbf{d}) TDDFT result for the time-derivative of the electric current yielding HH7, for $\epsilon = 0.3$ and $\theta = \Gamma\text{X}+5^{\circ}$. The red curves show the
    $x$ and $y$ projections of the driving laser field.
    }
\end{figure}
Three scenarios are in principle possible to explain the observation of circular harmonics from elliptical driver pulses shown in Figs. \ref{ellipdriver} and \ref{CompExpTDDFT}:
First, the harmonic emission occurs directly with this polarization state. Second, the harmonics are emitted with elliptical polarization and subsequently changed during propagation. Third, the driving pulse's polarization is altered during propagation due to induced birefringence. Moreover, the presence of the surface and a possible oxide layer might affect the polarization of the harmonics.

To address this question, we performed extensive microscopic TDDFT simulations (see Methods section), which at this point do neither account for propagation nor surface effects, computing only the nonlinear microscopic response of the crystal to the incident electric field.  For varying $\epsilon$ and $\theta = \Gamma\text{X}$, we computed \textit{ab initio} the high-harmonic response from Si and compared it to our measurements.  The results shown in Fig.~\ref{CompExpTDDFT}a-c display a remarkable agreement between experimental data and TDDFT calculations. This is true for harmonic yield, harmonic ellipticity as well as the  rotation of the harmonics' major axes.  We find minor deviations between calculations and experiments, mostly for HH7 and HH9, which can be expected by the increasing role of light-propagation effects for photon energies above the bandgap. However, even in presence of a surface and propagation effects in experiment, the calculations yield circular harmonics from elliptical drivers exactly for the conditions in which they are observed experimentally. This is shown for HH7 in Fig. \ref{CompExpTDDFT}d.
Therefore our {\it ab-initio} simulations confirm unambiguously that the measured polarization states of the harmonics have a microscopic origin in the coupled inter- and intraband dynamics, and is not due to macroscopic propagation effects or induced birefringence. From the comparison between experiments and simulations, it seems that the surface does not play a major role in determining the polarization states of the emitted harmonics.

In conclusion, after the first works on circular HHG from solids \cite{nicolasncomm17,Klemke_Atto17,Saito_Optica17}, we aimed at advancing our understanding and to demonstrate that a high degree of control over the polarization states of HHG from solids can be achieved. We found that both crystal symmetry and the nonperturbative coupled interband and intraband dynamics underlying harmonic emission play decisive roles in the polarization states of the emitted harmonics.
We have elucidated this duality between symmetry and dynamics in experiments on high-harmonic generation from silicon and quartz accompanied by {\it ab-initio} TDDFT simulations.
Our investigation has revealed that both the yields and polarization states of the higher harmonics sensitively respond differently to driver pulse ellipticity, sample rotation, and intensity.
In a broader perspective, our results demonstrate that the relative importance of intraband and interband mechanisms is not only determined by the driving wavelength and the material itself, but can be dynamically controlled by the laser intensity.

Circular harmonics can be produced for both circular and elliptical driver polarizations:
For circular driver pulses, the circular harmonics have alternating helicities, consistent with the selection rules derived from the crystallographic
point-group symmetry \cite{Tang71}.
For elliptical driver pulses, circular harmonics were generated for the first time to our knowledge, with up to 40\% efficiency compared to linear driver pulses in Si, corresponding to an 18$\times$ enhancement compared to circular harmonics from circular drivers. %
Compact sources of bright circular harmonics from solids extending into the XUV regime might open up appealing new applications in the spectroscopy of chiral systems, magnetic materials, and 2D materials with valley selectivity \cite{Liu_NP17}.
Circular isolated attosecond pulses from solids also seem in reach employing appropriate gating techniques.
Finally, polarization-state-resolved high-harmonic spectroscopy offers the unique advantage of sensitivity to both electronic and structural dynamics on sub-cycle time scales, thus opening up new avenues for the spectroscopy of quantum materials on extreme time scales
\cite{BasovNM17,NicolasNIO17}.

\subsection*{Methods}

\subsection*{Experimental high-harmonic generation setup}
\noindent Supplementary Fig. 1 shows the experimental setup used for high-order harmonic generation (HHG) from crystalline solids.
Passively carrier-envelope phase (CEP)-stabilized \cite{Cerullo_LPR11}, 120-fs pulses at 2.1 $\mu$m (0.59 eV photon energy) are generated in a Ti:sapphire-pumped white-light-seeded
optical parametric amplifier (OPA) \cite{Rossi_OL18,Muecke_IEEE15}.
These 2.1-$\mu$m driver pulses pass through a wire-grid polarizer, a quarter-wave plate (QWP) and a half-wave plate (HWP),
which allow setting the driver ellipticity while keeping the major axis of the polarization ellipse constant (see Supplementary Fig. 2).
The pulses are focused onto the
sample with a 25-cm CaF$_2$ lens, resulting in a $1/e^2$ focus diameter of $2w_0=95\,\mu$m.
After $50\,$cm of propagation, an iris is used to spatially suppress the otherwise very strong third harmonic. A curved UV-enhanced Al mirror is used to direct the output light to an Ocean Optics UV-VIS HR4000 spectrometer with a slit width of 10 $\mu$m. To determine the ellipticities and major axes of the generated harmonics, a Rochon polarizer is placed between sample and iris and rotated in total by $360^{\circ}$, measuring a spectrum every $18^{\circ}$. To detect the helicity of the circular harmonics, a tunable zero-order QWP (from Alphalas) is placed between sample and polarizer.
For post-processing the polarizer scans, the harmonic intensities are fitted with a sin-square curve offset from zero, the ellipticity calculated as
$|\epsilon_{\rm HH}| = \sqrt{I_{\rm min}/I_{\rm max}}$ and the major-axis rotation as $\phi_{\rm HH} = \arctan(I_{\rm y}/I_{\rm x})$. The driving-intensity scan in Supplementary Fig. 3 is performed employing reflective neutral-density filters.

\subsection*{\emph{Ab-initio} TDDFT simulations of high-harmonic generation in solids}

\noindent Within the framework of time-dependent density functional theory (TDDFT), the evolution of the wavefunctions and the evaluation of the time-dependent current are computed by propagating the Kohn-Sham equations
\begin{equation}
 i\hbar\frac{\partial \psi_{n,\mathbf{k}}(\mathbf{r},t)}{\partial t} = H_{\rm KS}(\mathbf{r},t)\psi_{n,\mathbf{k}}(\mathbf{r},t),
\end{equation}
where $\psi_{n,\mathbf{k}}$ is a Bloch state, $n$ a band index, $\mathbf{k}$ a point in the first Brillouin zone (BZ), and $H_{\rm KS}$ is the Kohn-Sham Hamiltonian given by
\begin{equation}
 H_{\rm KS}(\mathbf{r},t) =\frac{1}{2m}(-i\hbar\nabla +\frac{e}{c}\mathbf{A}(t))^2 +v_{\mathrm{ext}}(\mathbf{r},t)+v_{\mathrm{H}}(\mathbf{r},t)+v_{\mathrm{xc}}(\mathbf{r},t)\,.
\end{equation}
The different terms correspond to the kinetic energy, the ionic potential, the Hartree potential, that describes the classical electron-electron interaction, and exchange-correlation potential, that contains all the correlations and nontrivial interactions between the electrons. The latter needs to be approximated in practice \cite{Rubio_TDDFT_book}.

We perform the calculations using the Octopus code~\cite{C5CP00351B}, employing the TB09~\cite{PhysRevLett.102.226401} meta generalized gradient approximation (MGGA) functional to approximate the exchange-correlation potential using the adiabatic approximation. We employ norm-conserving pseudo-potentials.
We emphasize that within TB09 MGGA, the experimental band gap of common semiconductors and insulators is well reproduced \cite{PhysRevB.87.075121}, which is an important improvement over the local-density approximation (LDA) used in [\onlinecite{nicolasPRL},\onlinecite{nicolasncomm17}], permitting direct comparison between experiment and theory.
As shown in Ref. [\onlinecite{nicolasPRL}], dynamical correlations do not affect the HHG spectra of Si. The excitonic effects in Si mainly come from the long-range part of the exchange-correlation potential~\cite{botti2004long}, i.e., a renormalization of the Hartree term (which does not play any role in HHG from Si~\cite{nicolasPRL}), therefore excitonic effects are not expected to modify the HHG spectra of materials such as Si. We note that this is not necessarily true for all materials, in particular materials with strongly localized excitons, for which bound states will form in the band gap, or in strongly correlated materials \cite{NicolasNIO17,Silva_NP18}.

All calculations for bulk Si are performed using the primitive cell of bulk Si, using a real-space spacing of 0.484 atomic units. We consider a laser pulse of 50-fs FWHM duration with a sin-square envelope and a carrier wavelength $\lambda$ of 2.08\,$\mu$m, corresponding to 0.60\,eV carrier photon energy.
We employ an optimized 36$\times$36$\times$36 grid shifted four times to sample the BZ, and we use the intensity corresponding to the experimental intensity, using the value for the optical index $n$ of $\sim 3.4$ for computing the intensity in matter.
We use the experimental lattice constant $a$ leading to a MGGA band gap (direct) of silicon of 3.09\,eV. In all our calculations, we assume a CEP of $\phi=0$.

We compute the total electronic current  $\mathbf{j}(\mathbf{r},t)$ from the time-evolved wavefunctions, the HHG spectrum is then directly given by
\begin{equation}
 \mathrm{HHG}(\omega) = \left|\mathrm{FT}\Big(\frac{\partial}{\partial t}\int d^3\mathbf{r}\, \mathbf{j}(\mathbf{r},t)\Big)\right|^2,
\end{equation}
where $\mathrm{FT}$ denotes the Fourier transform.

Supplementary Fig. 5 shows a comparison of a computed HHG spectrum to a corresponding experimental spectrum. Note that, as mentioned above, in our experiments we only detect harmonics up to HH9 due to the spectrometer used (Ocean Optics UV-VIS HR4000). Our TDDFT calculations predict that harmonics up to  HH19 in the XUV spectral region are generated for our experimental conditions.

\subsection*{Data availability}
\noindent The data that support the findings of this study are available from the corresponding authors on reasonable request, and will be deposited on the NoMaD repository (https://nomad-repository.eu/).

\subsection*{Code availability}
\noindent The OCTOPUS code is available from http://www.octopus-code.org.

\bibliographystyle{naturemag}

\subsection*{Acknowledgements}
\noindent We acknowledge support by the excellence cluster 'The Hamburg Centre of Ultrafast Imaging-Structure, Dynamics and Control of Matter at the Atomic Scale', the priority program QUTIF (SPP1840 SOLSTICE) of the Deutsche Forschungsgemeinschaft, and financial support from the European Research Council (ERC-2015-AdG-694097), Grupos Consolidados (IT578-13), and  European Union's H2020  program under GA no.676580 (NOMAD).
N.T.-D., A.R., and O.D.M. thank M. Altarelli for very fruitful discussion. We thank M. Spiwek for help with the Laue X-ray diffraction characterization of samples.

\subsection*{Author contributions}
\noindent N.T.-D., A.R., F.X.K. and O.D.M. conceived, designed and coordinated the project.
G.M.R. and R.E.M. implemented the IR-OPA driver source.
N.K., Y.Y., G.D.S. and O.D.M. conceived the setup and performed the HHG experiments.
N.T.-D. carried out the code implementation and numerical calculations.
N.K., N.T.-D. and O.D.M. analyzed and interpreted the experimental and theoretical results.
N.K., N.T.-D., A.R., F.X.K., and O.D.M. participated
in the discussion of the results and contributed to the manuscript with revisions by all.

\newpage

\makeatletter
\renewcommand*{\fnum@figure}{Supplementary Fig. {\normalfont \thefigure}}
\renewcommand*{\@caption@fignum@sep}{. }
\makeatother
\setcounter{figure}{0}

\section*{\large Supplementary information}

\section{Experimental high-harmonic generation setup}
\noindent Supplementary Fig. 1 shows the experimental setup used for high-order harmonic generation (HHG) from crystalline solids.
For a detailed discussion of this setup, see Methods section.
\begin{figure}[h!]
    \centering
    \includegraphics[width=\columnwidth]{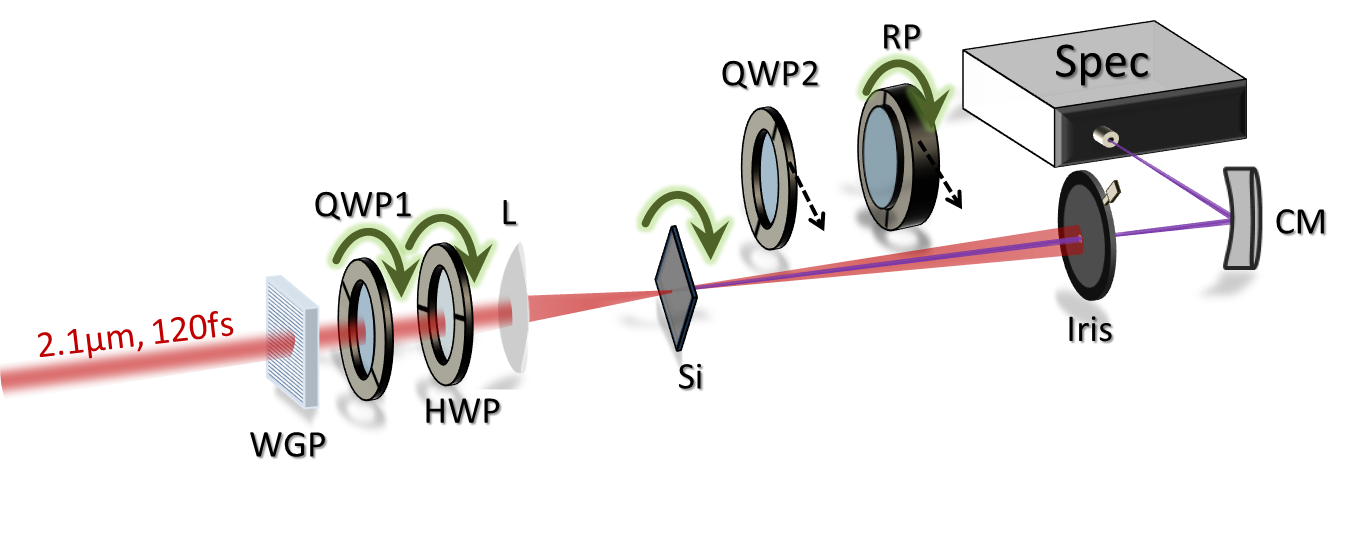}
    \vspace{-8mm}
    \caption{\textbf{Experimental HHG setup.} WGP, Wire-grid polarizer; QWP1, quarter-wave plate; HWP, half-wave plate; L, CaF$_2$ lens, $f=25$\,cm; Si, (100)-cut 2-$\mu$m-thin silicon; RP, Rochon polarizer; QWP2, tunable quarter-wave plate; CM, curved mirror; SPEC, UV-NIR spectrometer.
     }
\end{figure}

\section{Precision of ellipticity and major-axis angle calibration}

\noindent When measuring the harmonic response maps as function of driver-pulse ellipticity $\epsilon$ and sample rotation $\theta$,
we require that the major axis of the driving field polarization ellipse remains constant ideally over the entire range of $\epsilon$ values. Furthermore, also the precise value of $\epsilon$ is important.
As can be seen from the calibration data shown in Supplementary Fig. 2, the combination of a quarter-wave plate (QWP) and a half-wave plate (HWP) allows us to vary to ellipticity $\epsilon$ within [0, 0.98], while the major axis of the polarization ellipse remains constant to $\pm 2^{\circ}$ within $|\epsilon| < 0.8$ and $\pm 4^{\circ}$ for $|\epsilon| > 0.8$.
\begin{figure}[h!]
    \centering
    \includegraphics[width=0.6\columnwidth]{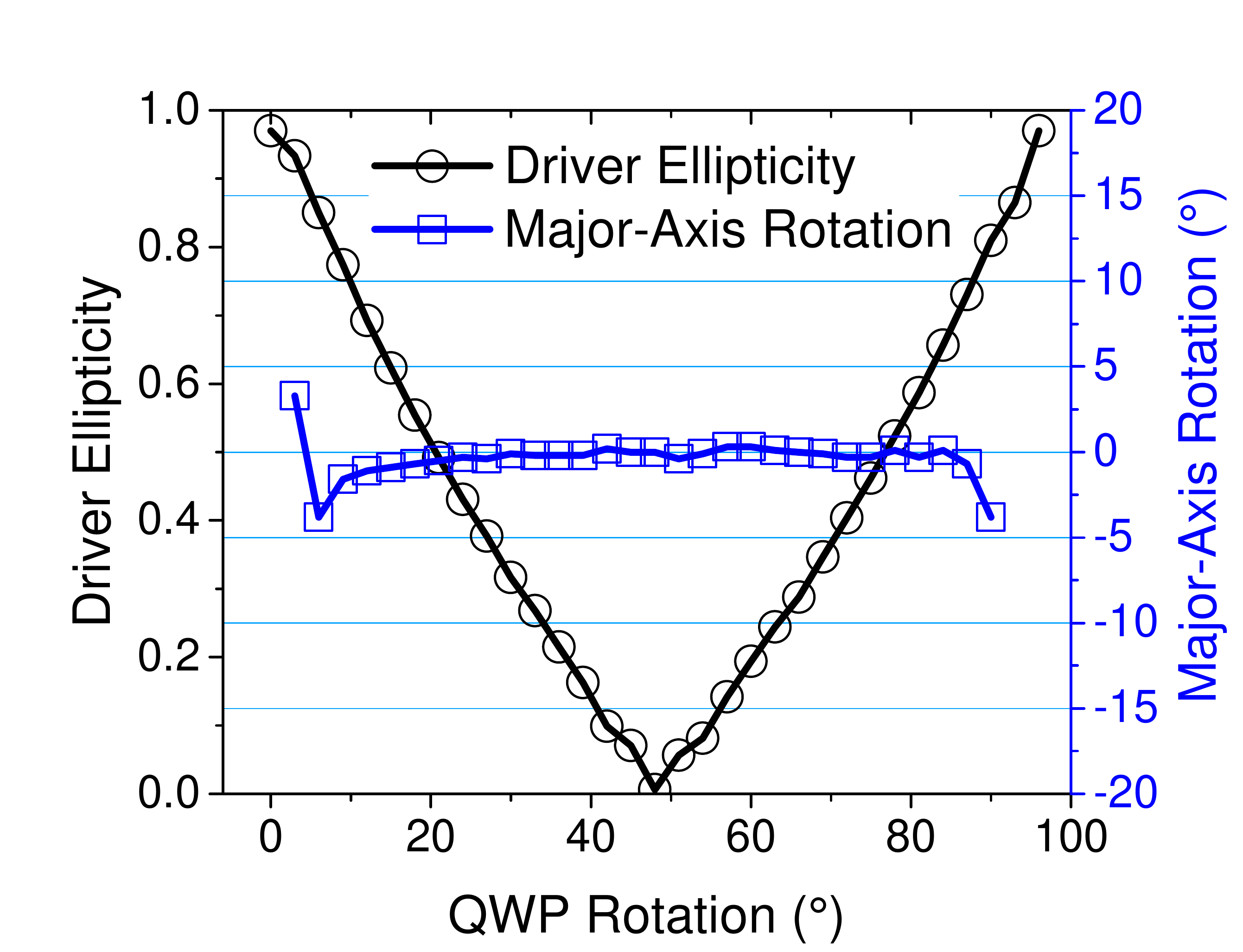}
      \vspace{-2mm}
    \caption{\textbf{Ellipticity and major-axis angle calibration.} The ellipticity and major-axis orientation of the driving field as measured with a polarizer and powermeter. The major axis is kept constant to $\pm 2^{\circ}$ within $|\epsilon| < 0.8$ and $\pm 4^{\circ}$ for $|\epsilon| > 0.8$.}
\end{figure}

\section{Intensity scaling of the harmonics in silicon}

\noindent Perturbatively, the $n^{\text{th}}$-harmonic intensity, $I_n$, scales as $I_n \sim I^n$. Here, $I$ is the driving pulse intensity. Supplementary Fig. 3 shows the intensities of HH5-HH9 in Si when varying the driving intensity. For intensities above $0.3~$TW\,cm$^{-2}$, all harmonics clearly deviate from the perturbative power law. All our experiments and simulations are performed above this intensity in the nonperturbative regime.
\begin{figure}[h!]
    \centering
    \vspace{-4mm}
    \includegraphics[width=0.7\columnwidth]{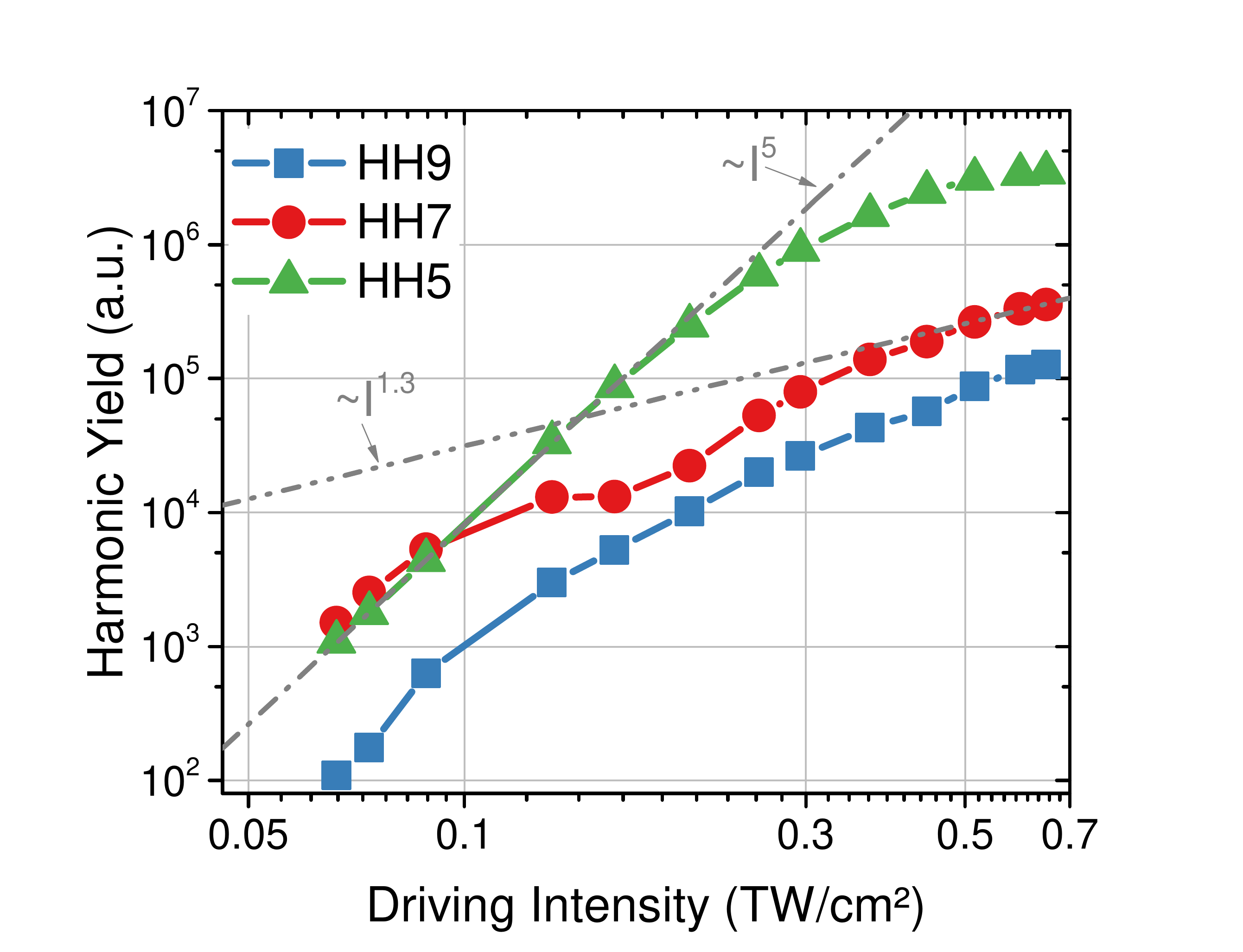}
    \vspace{-2mm}
    \caption{\textbf{Dependence of the measured harmonic yield on the driving intensity for \boldmath$\epsilon = 0$.}
     }
\end{figure}

\section{Laue X-ray diffraction of silicon samples}

\noindent In our experiments, we mainly use free-standing 2-$\mu$m-thin (100)-cut Si samples (by Norcada). To verify the monocrystallinity and to determine and align the exact crystal orientation,  in-house Laue X-ray diffraction is performed employing a Mo anode X-ray tube on these samples. Supplementary Fig. 4 shows a typical diffraction pattern.
\begin{figure}[h!]
    \centering
    \includegraphics[width=0.4\columnwidth]{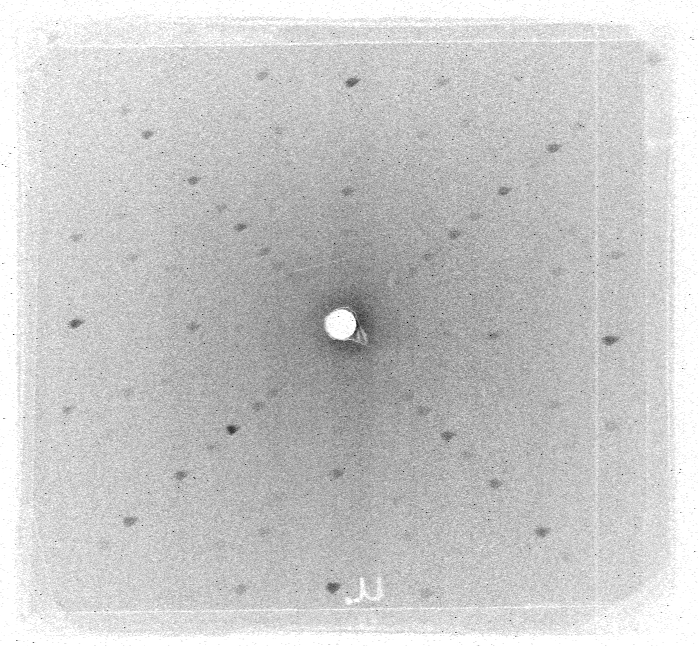}
    \caption{\textbf{Laue X-ray diffraction pattern from a (100)-cut Si sample.}\hspace{3cm}{}}    %
\end{figure}

\section{Comparison of TDDFT computed and experimentally measured HHG spectra}

\noindent Supplementary Fig. 5 shows a comparison of a computed HHG spectrum to a corresponding experimental spectrum. Note that in our experiments we only detect harmonics up to HH9 due to the spectrometer used (Ocean Optics UV-VIS HR4000). Our TDDFT calculations, described in the Methods section, predict that harmonics up to  HH19 in the XUV spectral region are generated for our experimental conditions.
\begin{figure}[h!]
    \centering
    \includegraphics[width=0.8\columnwidth]{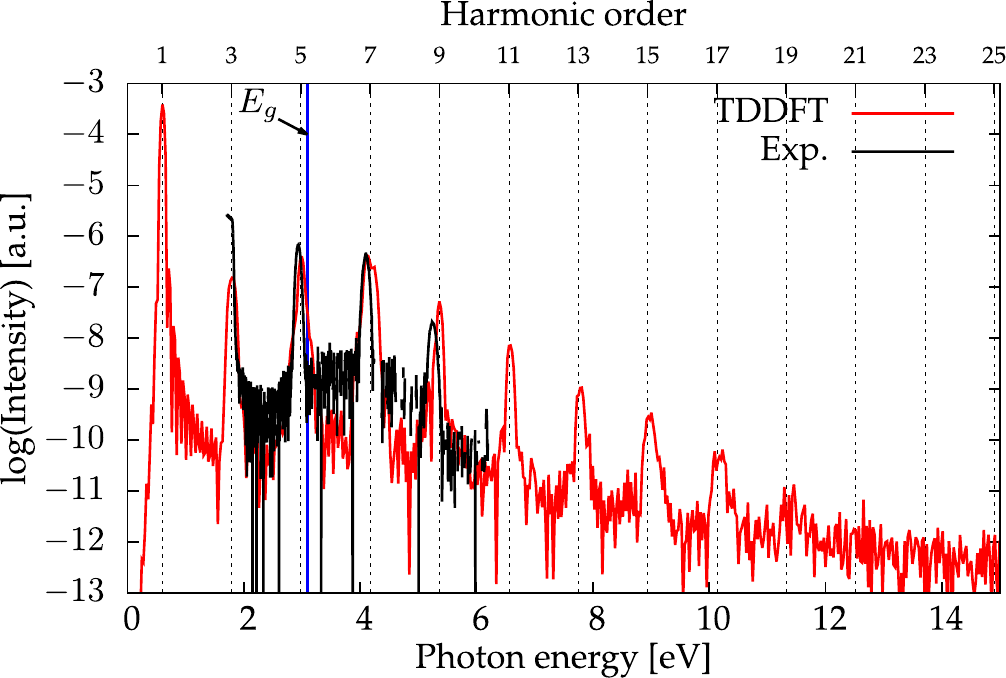}
    \vspace{-3mm}
    \caption{\textbf{Computed harmonic spectrum of Si compared to the experimental spectrum.} The driving intensity in vacuum is $6\times10^{11}\,$W\,cm$^{-2}$, driving wavelength is $\lambda=2.08$\,$\mu$m. The driving field is linearly polarized and the polarization direction is taken along the $\Gamma$X direction. The blue line indicates the theoretical band-gap calculation result using the TB09 MGGA functional. The experimental wavelength is 2.1\,$\mu$m.}
\end{figure}

\section{Joint density of states (JDOS)}

\noindent The generation of harmonic photons by the interband mechanism corresponds to the recombination of
an excited electron in a conduction band with a hole in a valence band. Therefore, the possibility of emitting a photon by the interband mechanism at a given energy is directly linked to the density of optical transitions available at this photon energy~\cite{nicolasPRL}. This quantity is also known as the joint density of states (JDOS).
Due to the nonperturbative nature of the process of high-harmonic generation, carriers intially created by interband transitions near the $\Gamma$ point in Si, start to explore a region of the BZ around the minimal band gap.
In order to estimate the region of the BZ seen by the driven electrons, one can resort to the so-called ``acceleration theorem'', which gives an upper limit to this region. In fact, the scattering between electrons, as well as the coupling to other bands (interband transitions) reduces the exploration of the electron wavepacket~\cite{nicolasncomm17}.
The JDOS is given by
 \begin{equation}
  \mathrm{JDOS}(\omega) \propto \sum_{\rm v,c} {\sum_{\mathbf{k}}}^{'} \delta(E_{{\rm c},\mathbf{k}}-E _{{\rm v},\mathbf{k}} -\omega)\,,
 \end{equation}
where the indices v,c denote valence and conduction bands, $\sum_{\mathbf{k}}^{'}$ denotes a sum over the explored region of the BZ determined by the acceleration theorem in the case of independent electrons, i.e., $\mathbf{k}(t) = -\frac{1}{c}\mathbf{A}(t)$.

In Supplementary Fig. 6, we show the evolution of the JDOS with the ellipticity of the driving field and sample rotation. We found that for all sample rotations and ellipticities, the JDOS at the energy of HH7 is much higher than at the energy of HH9, indicating that interband transitions are stronger for HH7 compared to HH9.
\begin{figure}[h!]
     \centering
     \vspace{-3mm}
     \includegraphics[width=\columnwidth]{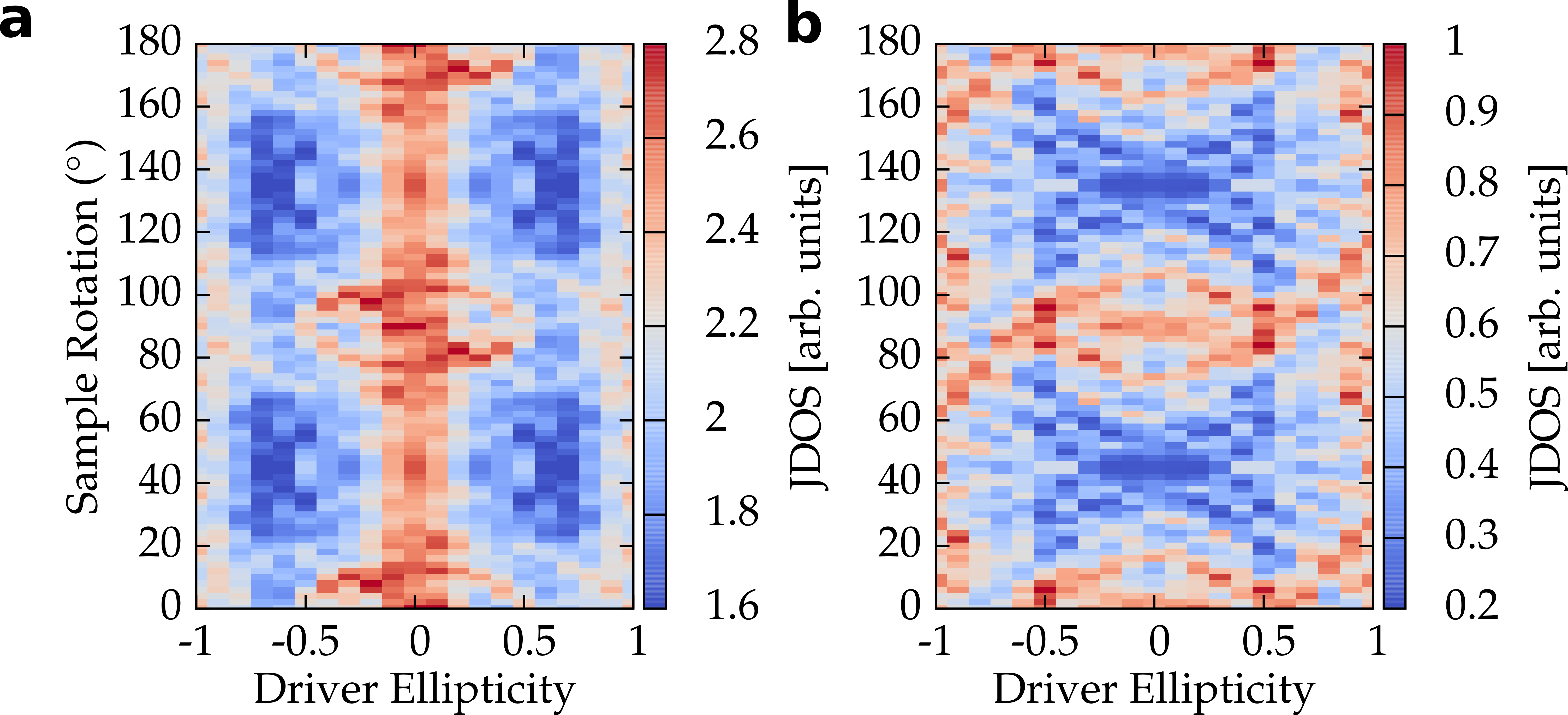}
     \vspace{-4mm}
     \caption{\textbf{Dependence of the JDOS on driver ellipticity and sample rotation.} JDOS at the energy of \textbf{(a)} HH7 and \textbf{(b)} HH9. HH5 is not shown as it lies below the experimental band gap of Si. Calculations are performed for an intensity in vacuum of 0.6\,TW\,cm$^{-2}$.}
 \end{figure}

\section{Calculation of the center of mass}

\begin{figure}[b]
    \centering
    \vspace{-3mm}
    \includegraphics[width=0.65\columnwidth]{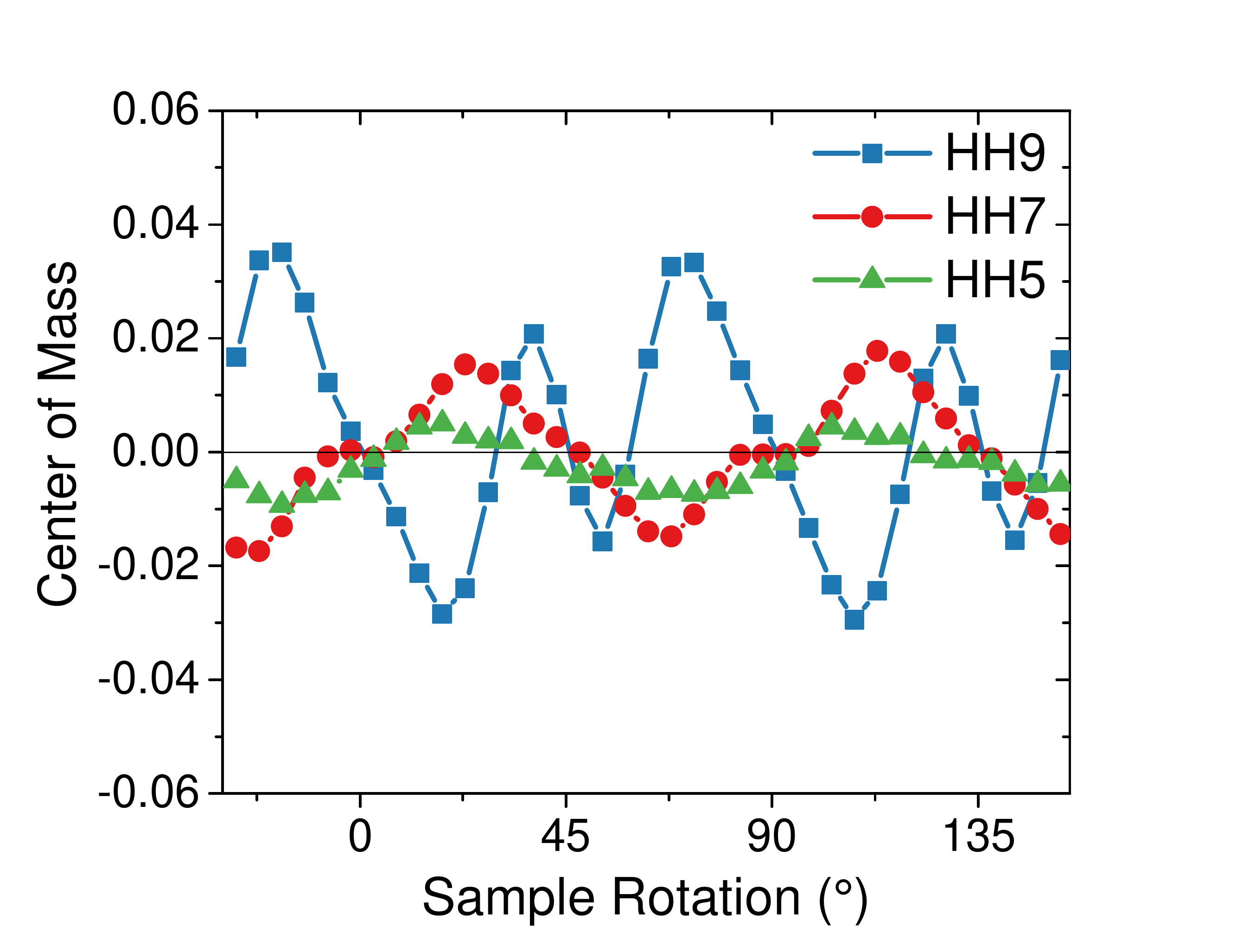}
     \vspace{-3mm}
    \caption{\textbf{Center of mass (COM) in $\epsilon$-coordinate for HH5, HH7 and HH9 in Si as shown in Fig. 1a-c in the main text.}}
\end{figure}
\noindent The center of mass (COM) curves displayed in Fig. 1a-c are calculated using
\begin{align}
{\rm COM}(\theta) = \frac{\sum\limits_{|\epsilon|<0.5} \epsilon \cdot I_{\rm HH}(\epsilon,\theta)}{\sum\limits_{|\epsilon|<0.5} I_{\rm HH}(\epsilon,\theta)},
\end{align}
where $I_{\rm HH}$ is the yield of the respective harmonic order, and $\epsilon$, $\theta$ are again driver ellipticity and sample rotation. The centers of mass are calculated in an interval $|\epsilon|<0.5$ to emphasize the asymmetric response in the most intense region of the ellipticity profiles. Supplementary Fig. 7 shows the COM curves of HH5-HH9 in Si as shown in Fig. 1a-c in the main text. Note that $\text{COM}=0$ is required by mirror symmetry along the symmetry axes $\Gamma$X and $\Gamma$K and that this is well reproduced in our data.

\section{Stokes polarization parameters of harmonics from silicon}

\noindent The polarization state of light can be fully characterized by the four Stokes parameters $S_0$, $S_1$, $S_2$ and $S_3$ [\onlinecite{Collett05},\onlinecite{Schaefer07}] defined via

\begin{align}
S_0 &= |E_x|^2 + |E_y|^2 \\
S_1 &= |E_x|^2 - |E_y|^2\\
S_2 &= 2\,{\rm Re}(E_xE_y^*)\\
S_3 &= -2\,{\rm Im}(E_xE_y^*)\,.
\end{align}

\noindent $E_x$ and $E_y$ denote the electric field components along the $x$- and $y$-directions. The degree of polarization (DOP) is defined as

\begin{align}
{\rm DOP} = \frac{\sqrt{S_1^2+S_2^2+S_3^2}}{S_0},\hspace{5mm}{\rm DOP}\in[0,1]\,.
\end{align}

\noindent ${\rm DOP}=1$ for completely polarized light, and ${\rm DOP}=0$ for completely unpolarized light.

If $I(\phi_{\rm pol},\delta_{xy})$ denotes the transmitted intensity after a polarizer with angle $\phi_{\rm pol}$ and an introduced phase shift $\delta_{xy}$ between $E_x$ and $E_y$, then the Stokes parameters can be determined from

\begin{align}
S_0 &= I(0,0) + I(\pi/2,0) \\
S_1 &= I(0,0) - I(\pi/2,0)\\
S_2 &= 2I(\pi/4,0) - S_0\\
S_3 &= S_0 - 2I(\pi/4, \pi/2)\,.
\end{align}

\noindent Hence, $S_0$, $S_1$ and $S_2$ can be measured with a polarizer only. For measuring $S_3$, an additional QWP is required to introduce $\delta_{xy}=\pi/2$.
The normalized Stokes parameters ($S_1/S_0$, $S_2/S_0$, $S_3/S_0$) can be measured throughout the IR to XUV regions \cite{Koide91,MingChang_NP18}.

\begin{figure}[t]
    \centering
    \includegraphics[width=\columnwidth]{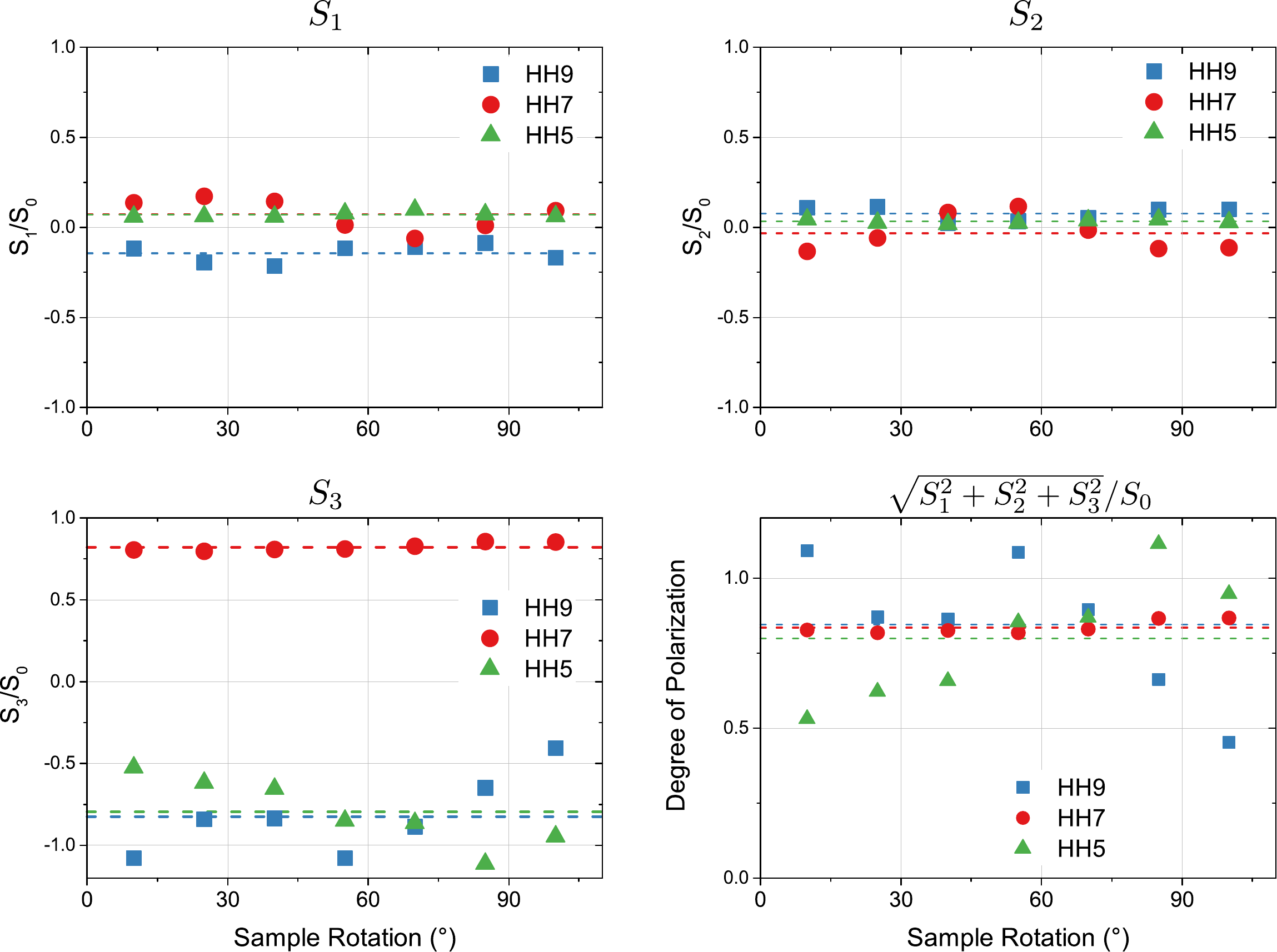}
    \caption{\textbf{Stokes parameters and degree of polarization versus sample rotation for circular driver polarization.} The dashed lines indicate the mean values.}
\end{figure}
\begin{figure}[b]
    \centering
    \includegraphics[width=\columnwidth]{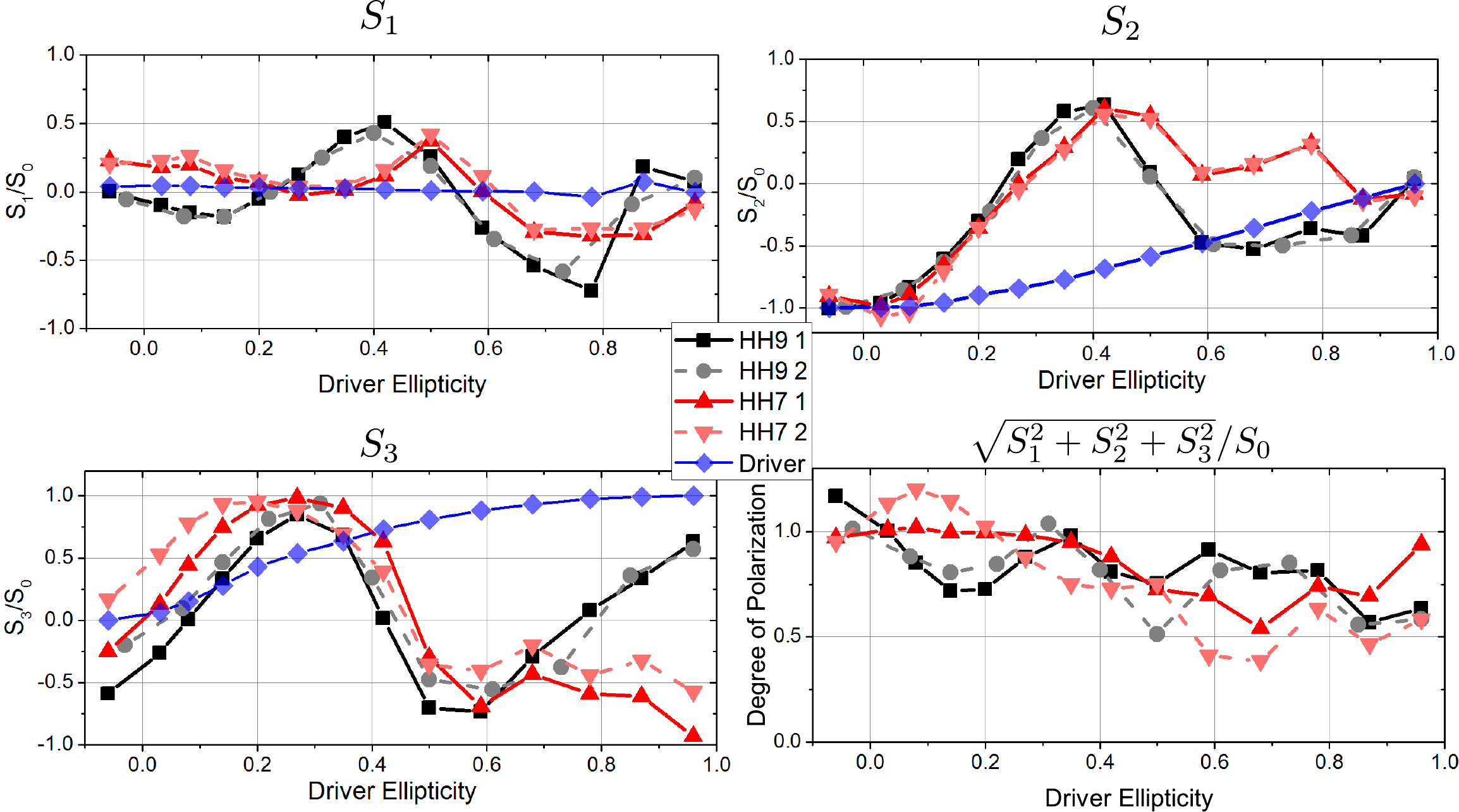}
    \caption{\textbf{Stokes parameters and degree of polarization versus driver ellipticity \boldmath$\epsilon$ for fixed sample orientation \boldmath$\theta = \Gamma\text{X}-5^{\circ}$.} Two independent measurements denoted by 1 and 2 are shown.}
\end{figure}
We measured the Stokes parameters of the harmonics from Si in two experiments:
for circular driver pulses versus sample rotation (see Supplementary Fig. 8), and for fixed sample orientation along $\theta=\Gamma\text{X}+5^{\circ}$ as function of driver ellipticity $\epsilon$ (Supplementary Fig. 9).
Note that the measurement of $S_3$ is known to be prone to experimental errors \cite{Schaefer07} due to the insertion of a QWP and comparison with another independent measurement without the QWP.
Our tunable QWP (from Alphalas) itself might absorb or reflect light and might also not introduce a precise $\pi/4$ phase shift for the different harmonics.
We believe that this is the reason for the scatter of $S_3$ values in Supplementary Fig. 8.
However, as stated in the main text and as can be seen from Supplementary Fig. 2, experimentally a circular polarization was only realized  to $\epsilon =0.98$,
therefore small variations of $S_3$ might also be caused by a minor deviation from perfectly circular driver polarization.

The different signs of $S_3$ of the individual harmonics observed in Supplementary Fig. 8 are another proof of circular harmonics with \emph{alternating} helicities. $S_1$ and $S_2$ are near zero, as one would expect for circular harmonics. The DOP values, that are comparable to reported values for the generation of circular harmonics from atomic and molecular gases \cite{Kfir2014,Veyrinas16}, show that the harmonics are highly polarized.

Supplementary Fig. 9 shows the evolution of the Stokes parameters of HH7 and HH9 when varying the driver-pulse ellipticity $\epsilon$, while keeping the major-axis orientation fixed.
In Supplementary Fig. 9, also the measured Stokes parameters of the driving field are displayed.
In order to also estimate the experimental uncertainty, we repeated the same measurements twice.
One can see clear deviations from the driver in the polarization-response of the various harmonics in all Stokes parameters $S_i$.
For $S_1$ and $S_2$, both measurements yield almost the same values. For $S_3$ and DOP, the insertion of the QWP introduces some uncertainty that, however, does not change the qualitative behavior.

\bibliographystyle{naturemag}

\end{document}